\documentclass[aps,showpacs,twocolumn]{revtex4}
\usepackage{amsfonts}
\usepackage{amsmath}
\usepackage{amssymb}
\usepackage{graphicx}

\begin{document}

\title{Breather-like solitons extracted from the Peregrine rogue wave}
\author{Guangye Yang$^{1,2}$, Yan Wang$^{1}$, Zhenyun Qin$^{3}$, Boris A.
Malomed$^{4}$, Dumitru Mihalache$^{5,6}$, and Lu Li$^{1}$}
\email{llz@sxu.edu.cn}
\affiliation{$^{1}$Institute of Theoretical Physics, Shanxi University, Taiyuan, Shanxi,
030006, China}
\affiliation{$^{2}$Department of Physics, Shanxi Medical University, Taiyuan, Shanxi,
030001, China\\
$^{3}$School of Mathematics, Key Laboratory of Mathematics for Nonlinear
Science and Shanghai Center for Mathematical Sciences, Fudan University,
Shanghai 200433, China}
\affiliation{$^{4}$Department of Physical Electronics, Faculty of Engineering, Tel Aviv
University, Tel Aviv 69978, Israel}
\affiliation{$^{5}$Horia Hulubei National Institute for Physics and Nuclear Engineering,
P.O.B. MG-6, RO-077125 Magurele, Romania}
\affiliation{$^{6}$Academy of Romanian Scientists, 54 Splaiul Independentei, RO-050094
Bucharest, Romania}

\begin{abstract}
Based on the Peregrine solution (PS) of the nonlinear Schr\"{o}dinger (NLS)
equation, the evolution of rational fraction pulses surrounded by zero
background is investigated. These pulses display the behavior of a
breather-like solitons. We study the generation and evolution of such
solitons extracted, by means of the spectral-filtering method, from the PS
in the model of the optical fiber with realistic values of coefficients
accounting for the anomalous dispersion, Kerr nonlinearity, and higher-order
effects. The results demonstrate that the breathing solitons stably
propagate in the fibers. Their robustness against small random perturbations
applied to the initial background is demonstrated too.
\end{abstract}

\maketitle

\section{Introduction}

The Peregrine solution (PS), which was discovered more than 30 years ago
\cite{Peregrine}, is one of the soliton solutions of the nonlinear Schr\"{o}%
dinger (NLS) equation existing on top of a finite continuous-wave (CW)
background. Recently, the PS has drawn renewed attention due to its unique
localization dynamics \cite{Zakharov,Shrira,Khawaja,Chen}. Particularly, it
has been extensively studied as a prototype of the oceanic rogue waves, and
is called, in this context, the Peregrine rogue wave. Actually, the PS is a
limit case of other NLS solitons existing on the finite background, \textit{%
viz}., the Kuznetsov-Ma soliton, localized in the transverse dimension \cite%
{Kuznetsov,Ma}, and of the Akhmediev breather, which is localized in the
longitudinal dimension \cite{Akhmediev}. It was found that a small
perturbation on top of the CW background may grow into the full Peregrine
rogue wave, pumped by the modulation (Benjamin-Feir)\ instability of the
background \cite{Kharif,Slunyaev,Ruban,Andonowati}. Experimental
observations concerning the formation and dynamics of the Peregrine rogue
wave in diverse physical media, such as optical fibers, water-wave tanks,
and plasmas, have been recently reported \cite{Kibler,Chabchoub,Bailung}.
Splitting of the Peregrine rogue wave under non-ideal initial conditions,
which exhibits its instability, has been demonstrated too \cite{Hammani}.

Another species of solutions of equations of the NLS type is represented by
breathers (\textit{breathing solitons}), which periodically oscillate in the
course of the evolution. A well-known example is the dispersion-managed
soliton, in which the periodic dilatation and compression is induced by the
alternating map of the anomalous and normal group velocity dispersion \cite%
{Smith,Nakazawa,Yu,Shapiro,book,DM-review}. Generally, breathers exist in
systems with periodically modulated parameters, such as dispersion,
gain/loss and nonlinearity coefficients, which can be used for amplification
and compression of solitons \cite{book,Serkin1,Serkin2,RuiyuHao}. The
concept of breathing solitons is also known in the theory of nonlinear
dissipative systems governed by generic complex Ginzburg-Landau equations,
in which self-sustained breathers are supported by the balance between the
dispersion and nonlinearity, concomitant with equilibrium between gain and
loss. Well-known examples of such dissipative breathers are provided,
chiefly, by laser cavities and similarly organized systems featuring the
combination of amplification and dissipation \cite%
{Petv,PhysicaD,Deissler,lasers,Tlidi,Akhmediev1,AK,Rosanov,Vladimir,Song}.
Breathers are known too in models of nonlocal nonlinear media \cite%
{Bezuhanov,Zhanghuafeng,Strinic,Aleksic,Petrovic}. Furthermore,
breather-type solitons have been predicted in a variety of other physically
relevant settings, including, \textit{inter alia}, discrete \cite{PRL86:2353}
and continuous \cite{PRL95:050403} models of Bose-Einstein condensates,
third-harmonic generation in nonlinear optics \cite{OC229:391},
reaction-diffusion systems \cite{PRE74:066201}, elastic rods with the
mechanical nonlinearity represented by a combination of quadratic and cubic
terms \cite{PSSB249:1386}, and dynamical models of galaxies \cite{astro}.

In this work, using the spectral-filtering method, we aim to study the
generation and propagation of breather-like solitons generated by the
Peregrine rogue waves in optical fibers with anomalous dispersion, Kerr
nonlinearity and high-order effects. The results help to understand unique
properties of the Peregrine rogue waves and design experimental generation
of the breathers in optical fibers. In this connection, it may be relevant
to mention recent works which demonstrated the generation of multiple
solitons, both freely moving ones and arrays in the form of the Newton's
cradle \cite{cradle} (see also Ref. \cite{Lu}), as a result of splitting of
a high-order NLS breather under the action of the third-order dispersion,
and emission of solitons by Airy waves in the nonlinear fiber \cite{Marom}.

The paper is organized as follows. In the next Section, we recapitulate the
Peregrine solution of the NLS equation. Then we present a rational fraction
pulse, and discuss its propagation properties that are similar to those of
breathers. In Sec. III, we investigate the generation, propagation, and
stability of such oscillating solitons in the nonlinear fiber. Conclusions
are summarized in Sec. IV.

\section{The Peregrine solution for NLS equation and the rational fraction
pulse}

\begin{figure}[tbp]
\centering\vspace{0.0cm} \includegraphics[width=8.5cm]{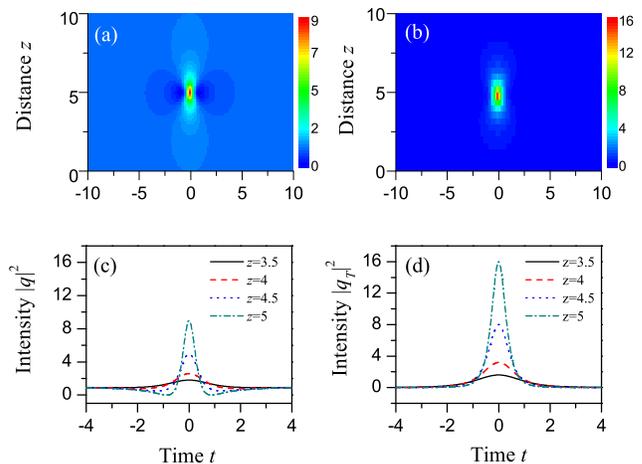} \vspace{0cm}
\caption{(Color online) The contour plots of (a) the Peregrine rogue wave
given by Eq. (\protect\ref{peregrine}) and (b) the rogue wave with zero CW
background given by Eq. (\protect\ref{newperegrine}). The growing intensity
distributions at propagation distances $z=3.5$, $z=4$, $z=4.5$, and $z=5$,
respectively, are shown in (c) for the Peregrine rogue wave, and in (d) for
the rogue wave with zero CW background. The parameters are $A=1$, $z_{0}=5$,
and $t_{0}=0$.}
\label{fig1}
\end{figure}

In the picosecond regime, the dynamics of the optical pulse propagation in
optical fibers is governed by the scaled NLS equation in the well-known form
\cite{Kodama,Agrawal},
\begin{equation}
i\frac{\partial q}{\partial z}+\frac{1}{2}\frac{\partial ^{2}q}{\partial
t^{2}}+|q|^{2}q=0,  \label{model}
\end{equation}%
where $q=q(z,t)$ is the slowly varying envelope of the electric field, $%
t=t^{\prime }-z/v_{g}$, $v_{g}$ stands for the group velocity, while $t$ and
$z$ represent the retarded temporal coordinate and normalized propagation
distance, respectively. The rational PS of the NLS equation is \cite%
{Peregrine}
\begin{equation}
q(z,t)=Ae^{iA^{2}Z}\left[ 1+R\left( Z,T\right) \right] ,  \label{peregrine}
\end{equation}%
with%
\begin{equation}
R(Z,T)=\frac{-4-i8A^{2}Z}{1+4A^{4}Z^{2}+4A^{2}T^{2}},
\end{equation}%
and $T=t-t_{0}$, $Z=z-z_{0}$, where $t_{0}$, $z_{0}$, and $A$ are arbitrary
real constants. Equation (\ref{peregrine}) demonstrates that the PS is a
superposition of a CW solution and a fraction function, in which the nonzero
CW background pushes an initial local weak pulse\ towards nonlinear
compression via the modulation instability. The largest compression of the
pulse is attained at $z=z_{0}$ ($Z=0$) \cite{Hammani}, as shown in Figs. \ref%
{fig1}(a) and \ref{fig1}(c). As a possible application, the maximally
compressed pulse can be used to generate high-power pulses, with the help of
the spectral-filtering technique \cite{Ygy1,Ygy2}. This method allows one to
eliminate the CW background around the maximally compressed pulse in the
temporal domain, therefore the shape of the high-power pulses, obtained by
means of the spectral filtering, is chiefly determined by the fraction part
of solution (\ref{peregrine}).

\begin{figure}[tbp]
\centering\vspace{0.0cm} \includegraphics[width=5.5cm]{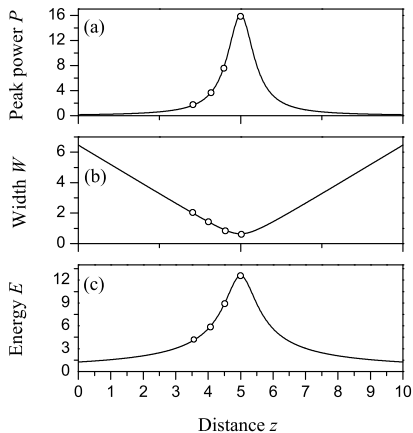} \vspace{0cm}
\caption{(Color online) The evolution of (a) the peak power, (b) FWHM, and
(c) energy of the rogue wave with zero background. The parameters are the
same as in Fig. \protect\ref{fig1}.}
\label{fig2}
\end{figure}

Now we turn to the consideration of the rational fraction part of solution (%
\ref{peregrine}), taking it as
\begin{equation}
q_{T}(z,t)=Ae^{iA^{2}Z}R\left( Z,T\right) .  \label{newperegrine}
\end{equation}%
This expression is \emph{not} an exact solution for the NLS equation, but it
is localized in time and the propagation direction, and is surrounded by
zero background, as seen in Fig. \ref{fig1}(b). Thus, for fixed $z$,
expression (\ref{newperegrine}) represents a fraction pulse, whose intensity
distributions at different distances are shown in Fig. \ref{fig1}(d).
Comparing Figs. \ref{fig1}(d) with Fig. \ref{fig1}(c), we concluded that the
corresponding intensities of the fraction pulse are larger than those of the
PS. Furthermore, it is straightforward to find the peak power, full-width at
half-maximum (FWHM), and energy of the rogue wave with zero background
\begin{align}
P\left( Z\right) & \equiv \left\vert q_{T}\left( z,t_{0}\right) \right\vert
^{2}=\frac{16A^{2}}{1+4A^{4}Z^{2}},  \notag \\
W\left( Z\right) & =\frac{\sqrt{(\sqrt{2}-1)(1+4A^{4}Z^{2})}}{A},  \notag \\
E\left( Z\right) & \equiv \int_{-\infty }^{+\infty }\left\vert
q_{T}\right\vert ^{2}dt=\frac{4A\pi }{\sqrt{1+4A^{4}Z^{2}}},  \label{PWE}
\end{align}%
as functions of propagation distance $z$. Figure \ref{fig2} shows their
evolution along $z$ (recall that $Z=z-z_{0}$). From here, one can see that,
at $z\leq z_{0}$, the peak power and energy of the fraction pulse increases
with the propagation distance $z$. One can also find that the peak power and
energy of this pulse increase with the decrease of the pulse's width until $%
z=z_{0}$, and the largest peak power and energy are attained, naturally, at
the narrowest pulse width. Small circles in Figs. \ref{fig2}(a-c) correspond
to the peak power, width and energy of the growing fraction pulse at $z=3.5$%
, $4$, $4.5$\ and $5$, see Fig. \ref{fig1}(d). Thus, the fraction pulse can
be used to generate narrow waveforms with prescribed peak powers and widths.

Next, we aim to find out whether the fraction pulse can generate a soliton
in the optical fiber \cite{Mahnke}. For this purpose, we consider the
dimensionless soliton number
\begin{equation}
N=\frac{W\sqrt{P}}{2\ln (1+\sqrt{2})},  \label{N}
\end{equation}%
where $P$ and $W$ are the initial peak power and FWHM for a sech-type pulse
\cite{Agrawal}. According to the standard soliton theory, any input pulse
tends to eventually transform into a fundamental soliton if the soliton
number falls into the interval of $0.5<N<1.5$. The breather wave form
appears at $N\neq 1$. For the fraction pulse given by Eq. (\ref{newperegrine}%
), it follows from expression (\ref{PWE}) that its energy is $E\left(
Z\right) =[\pi /(4\sqrt{\sqrt{2}-1})]P(Z)W(Z)$. It differs from energy $%
PW/\ln (1+\sqrt{2})$ of the sech-type pulse by factor $\pi \left( 4\sqrt{%
\sqrt{2}-1}\right) ^{-1}\ln \left( 1+\sqrt{2}\right) \approx 1.076$. As this
factor is close to $1$, the soliton number for this pulse can be
approximately identified according to Eq. (\ref{N}), which yields
\begin{equation}
N=\frac{W(Z)\sqrt{P(Z)}}{2\ln \left( 1+\sqrt{2}\right) }=\frac{\sqrt{4(\sqrt{%
2}-1)}}{\ln \left( 1+\sqrt{2}\right) }\approx 1.4604,  \label{soliton order}
\end{equation}%
where $P(Z)$ and $W(Z)$ are given by Eq. (\ref{PWE}). Thus, the emergence of
a breather oscillating around a fundamental soliton is expected.

\begin{figure}[tbp]
\centering\vspace{0.0cm} \includegraphics[width=8.5cm]{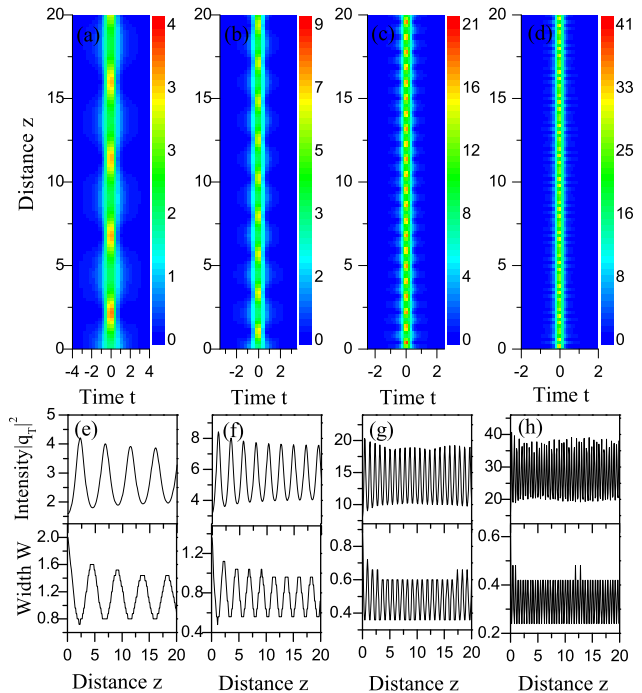} \vspace{0cm}
\caption{(Color online) The evolution of the rational fraction pulses,
initiated by the configurations extracted from Eq. (\protect\ref%
{newperegrine}) at the corresponding positions: (a) $z_{0}=3.5$; (b) $%
z_{0}=4 $; (c) $z_{0}=4.5$; (d) $z_{0}=5$ (see further details in the text).
(e-h) The evolution of the corresponding pulse intensity and FWHM. The
parameters are the same as in Fig. \protect\ref{fig1}.}
\label{fig3}
\end{figure}

We take the fraction pulse at different positions $z\equiv z_{0}$ as the
initial configuration, and simulate its subsequent evolution by numerically
solving Eq. (\ref{model}). The results are summarized in Fig. \ref{fig3},
which displays the subsequent evolution of the fraction pulse, initiated by
its configurations which are taken, as said here, at different positions
corresponding to $z_{0}=3.5$, $4$, $4.5$, and $5$. In Figs. \ref{fig3}(a)
through \ref{fig3}(d), the pulses exhibit a typical oscillatory behavior of
breathing solitons. With the increase of value $z_{0}$, at which the input
is extracted from the fraction pulse (\ref{newperegrine}), the oscillation
period and width of the generated breather become smaller, while its power
increases, as seen in Figs. \ref{fig3}(e) to \ref{fig3}(h). Thus, it is
possible to create high-power pulses with different widths and peak powers
by choosing appropriate input forms of the fraction pulse.

\section{Breather-like soliton generated by the spectral filtering}

The above results, obtained for the completely integrable NLS equation, are
relevant for picosecond pulses. In the femtosecond regime, higher-order
effects, such as the third-order dispersion, self-steepening, self-frequency
shift, and others, should be added to the model, transforming it into the
higher-order NLS equation \cite{Kodama,Agrawal}:
\begin{gather}
\frac{\partial A}{\partial \xi }+i\frac{\beta _{2}}{2}\frac{\partial ^{2}A}{%
\partial \tau ^{2}}-\frac{\beta _{3}}{6}\frac{\partial ^{3}A}{\partial \tau
^{3}}  \notag \\
=i\gamma \left( |A|^{2}A+\frac{i}{\omega _{0}}\frac{\partial (|A|^{2}A)}{%
\partial \tau }-T_{R}A\frac{\partial |A|^{2}}{\partial \tau }\right) ,
\label{HNLSE}
\end{gather}%
where $A(\xi ,\tau )$ is the slowly varying envelope of the electric field, $%
\tau $ and $\xi $ are the temporal coordinate and propagation distance, $%
\beta _{2}$ is the group-velocity dispersion, $\beta _{3}$ is the
third-order dispersion, $\gamma $ is the Kerr nonlinearity coefficient of
the fiber, $\omega _{0}$ is the central carrier frequency of the optical
field, and $T_{R}$ is the Raman time constant. Here, we take realistic
parameters for carrier wavelength $\lambda =1550$ nm, $\beta
_{2}=-8.85\times 10^{-1}$ ps$^{2}$/km, $\beta _{3}=1.331\times 10^{-2}$ ps$%
^{3}$/km, $\gamma =10$ W$^{-1}\cdot ~$km$^{-1}$ \cite{Kibler}, and $T_{R}=3$
fs, if the self-frequency shift is also considered.

By means of normalizations $A(\xi ,\tau )=\sqrt{P_{0}}q(z,t)$, $t=\tau
/T_{0} $, and $z=\xi /L_{D}$, with temporal scale $T_{0}=[\left\vert \beta
_{2}\right\vert /(\gamma P_{0})]^{1/2}$ and dispersion length $L_{D}=(\gamma
P_{0})^{-1}$, where $P_{0}$ is the initial power, Eq. (\ref{HNLSE}) can be
rewritten in the form
\begin{equation}
i\frac{\partial q}{\partial z}+\frac{1}{2}\frac{\partial ^{2}q}{\partial
t^{2}}+|q|^{2}q=i\alpha _{3}\frac{\partial ^{3}q}{\partial t^{3}}+i\alpha
_{4}\frac{\partial (|q|^{2}q)}{\partial t}+\alpha _{5}q\frac{\partial |q|^{2}%
}{\partial t},  \label{HNLSE1}
\end{equation}%
where $\alpha _{3}=\beta _{3}/(6\left\vert \beta _{2}\right\vert T_{0})$, $%
\alpha _{4}=-(\omega _{0}T_{0})^{-1}$ and $\alpha _{5}=T_{R}/T_{0}$. When
the higher-order terms in the right-hand side of Eq. (\ref{HNLSE1}) are
absent, Eq. (\ref{HNLSE1}) reduces to Eq. (\ref{model}). Here we take a
typical value of the initial power, $P_{0}=0.3$ W, hence the other
parameters are $T_{0}=0.543\,14$ ps, $L_{D}=1/3$ km, $\alpha
_{3}=0.004\allowbreak 615$, $\alpha _{4}=-0.001\allowbreak 514$, and $\alpha
_{5}=0.005523\allowbreak 4$.

As discussed in Ref. \cite{Ygy}, the PS can be excited by a small localized
(single-peak) perturbation pulse placed on top of a CW background. Here, as
a typical example, we take the initial condition as a Gaussian-shaped
perturbation pulse on the CW background
\begin{equation}
A(0,\tau )=\sqrt{P_{0}}[1+\epsilon \exp (-\tau ^{2}/2T_{1}^{2})],
\label{initial condition}
\end{equation}%
with amplitude $\epsilon $ and width $T_{1}$.

\begin{figure}[tbp]
\centering\vspace{0.0cm} \includegraphics[width=8.5cm]{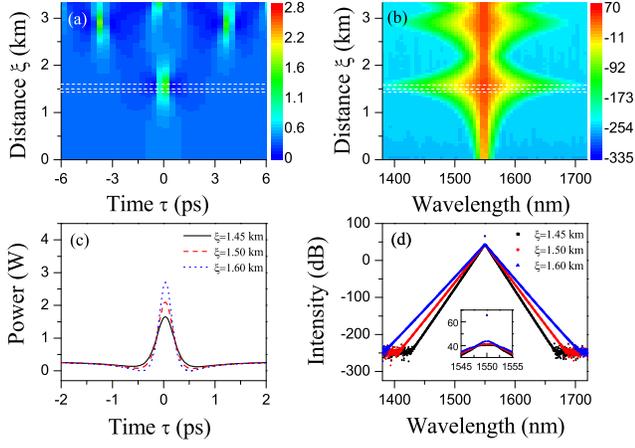} \vspace{0cm}
\caption{(Color online) The evolution of (a) the Peregrine rogue wave
excited by initial condition (\protect\ref{initial condition}), and (b) the
corresponding evolution of its spectral intensity. (c) The growing
temporal-domain pulse shapes and (d) the corresponding spectral shapes at $%
\protect\xi =1.45$ km, $1.50$ km, and $1.60$ km, respectively. The
parameters are $P_{0}=0.3$ W, $\protect\epsilon =0.0765$, and $T_{1}=3.7722$
ps.}
\label{fig4}
\end{figure}

In the absence of the Raman effect, the evolution of the initial
configuration (\ref{initial condition}) is presented in Fig. 4. The results
show that the Gaussian perturbation on top of the CW background can excite a
non-ideal Peregrine rogue wave, eventually evolving into the pulse-splitting
regime due to the modulational instability of the CW background, as shown in
Fig. \ref{fig4}(a). Figure \ref{fig4}(b) shows the corresponding evolution
of the spectrum, which starts with narrow spectral components and then
spreads into a triangular-type shape, similar to the so-called
supercontinuum \cite{super}. Then, the spectrum shrinks and spreads in the
course of the propagation, but it does not recover the initial shape, due to
the occurrence of the pulse splitting. This means that the Peregrine rogue
wave cannot travel directly with a preserved shape, hence so it cannot be
used for the design of a robust transmission scheme in optical fibers. For
our choice of the parameters (which, as said above, are realistic for the
applications), the largest compression of the pulse is attained at $\xi
=1.60 $ km, as shown in Fig. \ref{fig4}(c). The pulse's intensity
distributions at $\xi =1.45$ km, $1.50$ km, and $1.60$ km are presented in
Fig. \ref{fig4}(c), and the corresponding spectral intensities are shown in
Fig. \ref{fig4}(d), which shows that the spectrum of the background is
mainly concentrated at $\lambda =1550$ nm.

\begin{figure}[tbp]
\centering\vspace{0.0cm} \includegraphics[width=8.5cm]{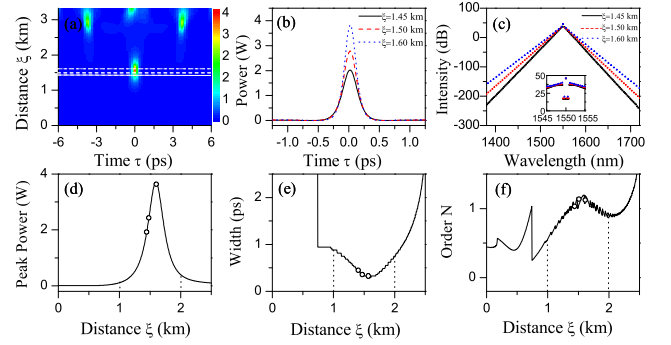} \vspace{0cm}
\caption{(Color online) (a) The evolution of the fraction pulse in the case
when the background was removed from the Peregrine rogue wave. (b) and (c)
The corresponding temporal and spectral shapes of the pulse at $\protect\xi %
=1.45$ km, $1.50$ km, and $1.60$ km, respectively. (d), (e), and (f) The
evolution of the peak power, FWHM, and soliton number $N$, respectively. The
parameters are the same as in Fig. \protect\ref{fig4}.}
\label{fig5}
\end{figure}

\begin{figure}[tbp]
\centering\vspace{0.0cm} \includegraphics[width=8.0cm]{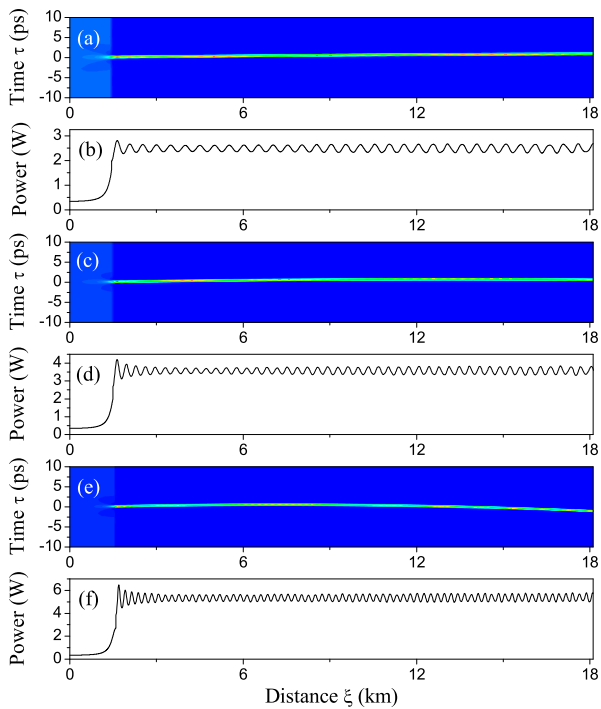} \vspace{0cm}
\caption{(Color online) (a), (c), and (e) The evolution of the initial state
(\protect\ref{initial condition}), corresponding to the fraction pulse, with
the background extracted from the PS at (a) $\protect\xi =1.45$ km, (c) $%
\protect\xi =1.50$ km, and (e) $\protect\xi =1.60$ km, respectively. (b),
(d), and (f) The corresponding evolution of the peak power. Parameters of
initial condition (\protect\ref{initial condition}) are the same as in Fig.
\protect\ref{fig4}.}
\label{fig6}
\end{figure}

\begin{figure}[tbp]
\centering\vspace{0.0cm} \includegraphics[width=8.0cm]{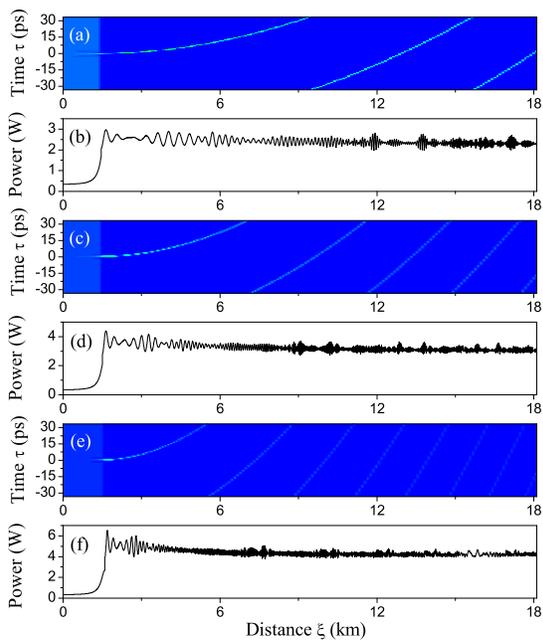} \vspace{0cm}
\caption{(Color online) The same as in Fig. \protect\ref{fig6}, but with the
Raman effect, with $T_{R}=3$ fs, taken into regard. The other parameters are
the same as in Fig. \protect\ref{fig4}.}
\label{fig7}
\end{figure}

\begin{figure}[tbp]
\centering\vspace{0.0cm} \includegraphics[width=8.0cm]{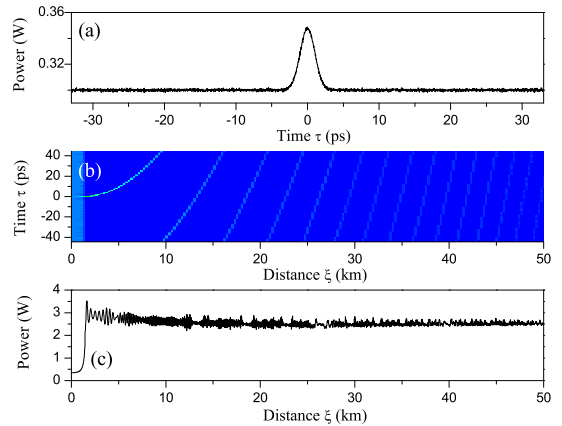} \vspace{0cm}
\caption{(Color online) The evolution of the breathing soliton extracted
from the PS at $\protect\xi =1.45$ km, under the the action of small
perturbations added to initial condition (\protect\ref{initial condition}).
(a) The initial profile (\protect\ref{initial condition}), perturbed by $%
0.1\%$ random noise. (b) The long-distance evolution of the pulse. (c) The
corresponding evolution of the peak power. The parameters are the same as in
Fig. \protect\ref{fig4}.}
\label{fig8}
\end{figure}

Thus, to use the PS for the generation of high-power pulses with
well-defined shapes, one needs to eliminate the background. Several ways for
this, such as the use of the polarization technique, Raman effect, and
interferometry, were previously proposed \cite{Stolen,Mamyshev,Fatome,Mahnke}%
. Here, we employ the spectral-filtering method to remove the background, as
discussed in Ref. \cite{Ygy1}, in which the spectrum around $\lambda =1550$
nm was filtered out by attenuating its spectral intensity by $10\%$. Figure %
\ref{fig5}(a) displays the evolution of the fraction pulse, in which the
background is filtered out in the range of $1$ nm around $1550$ nm. Such
pulses are indeed surrounded by zero background, as seen in Fig. \ref{fig5}%
(b). The corresponding temporal and spectral intensity shapes at $\xi =1.45$
km, $1.50$ km, and $1.60$ km, are displayed in Figs. \ref{fig5}(b) and \ref%
{fig5}(c), respectively. Comparing with Figs. \ref{fig4}(c) and \ref{fig4}%
(d), we conclude that the temporal intensities of the fraction pulse are
larger than those of the PS corresponding to the same propagation distances,
$\xi =1.45$ km, $1.50$ km, and $1.60$ km, and the pulse's spectrum is
attenuated around $1550$ nm. The peak power and FWHM of such pulses are
shown in Figs. \ref{fig5}(d) and \ref{fig5}(e). The respective soliton
number $N$, given by formula (\ref{soliton order}), is plotted in Fig. \ref%
{fig5}(f). Open circles in panels \ref{fig5}(d-f) correspond to the sequence
of growing pulses shown in Fig. \ref{fig5}(c). Note that the numerical
results are valid only in the interval of $1<\xi <2$, because the filtered
pulse does not exhibit a localized profile when the peak power is too small.
In Fig. \ref{fig5}(f) one can see that the corresponding soliton numbers are
all close to $N=1$ [see open circles in Fig. \ref{fig5}(f)], being obviously
smaller than the numbers for the picosecond pulse, as the width of the
pulses is smaller in the femtosecond regime.

Further, we have performed numerical simulations of the pulse evolution when
the background was filtered out from the PS at propagation distances $\xi
=1.45$ km, $\xi =1.50$ km, and $\xi =1.60$ km, respectively. The results are
summarized in Fig. \ref{fig6}, which demonstrates that the pulses propagate
stably over long propagation distances and exhibit the characteristic
behavior of the breathing soliton. In particular, typical oscillatory
evolution of the peak powers is shown in Figs. \ref{fig6}(b), \ref{fig6}(d),
and \ref{fig6}(f).

Next, we consider the influence of the self-frequency-shift effect on the
evolution of such pulses. The corresponding numerical results are displayed
in Fig. \ref{fig7}, which shows the typical evolution of the pulses in the
temporal domain, under the action of the Raman effect. Depending on the
input, the evolving breathing solitons become narrower, whereas the peak
powers increase accordingly, see Figs. \ref{fig7}(b), \ref{fig7}(d), and \ref%
{fig7}(f), respectively.

Finally, we consider the stability of the pulse against small random
perturbations added to the initial background. In the numerical simulations,
white-noise at the amplitude level of $0.1\%$ was added, as shown in Fig. %
\ref{fig8}(a). As a typical example, the evolution of the pulse which was
extracted from the PS at $\xi =1.45$ km is shown in Figs. \ref{fig8}(b) and %
\ref{fig8}(c). Although the background around the breather was effectively
removed at this point, the initial perturbations had a chance to contaminate
the main fraction pulse. The results reveal that the initial random
perturbations do not destabilize the generation and long-distance
propagation of the breathing soliton.

\section{Conclusions}

Based on the analytic solution of the nonlinear Schr\"{o}dinger equation in
the form of the PS\ (Peregrine soliton, alias the Peregrine rogue wave), the
evolution of a rational fraction pulse, extracted from that exact solution,
has been analyzed in detail. The extensive numerical results have shown that
the fraction pulse exhibits robust oscillatory (breather-like) propagation
in the realistic model of nonlinear dispersive optical fibers. We have also
investigated the generation and propagation of such breathing solitons
extracted from the PS with the Raman self-frequency shift also taken into
account. The numerical results clearly show that such solitons stably
propagate in the optical fibers, being stable against random perturbations.
The settings analyzed in this work can be experimentally implemented in the
nonlinear fibers.

\section{Acknowledgement}

This research is supported by the National Natural Science Foundation of
China grant 61078079 and 61475198,\ the Shanxi Scholarship Council of China
grant 2011-010 and the 331 Foundation of Basic Medical College of Shanxi
Medical University. Qin is sponsored by Shanghai Pujiang Program (No.
14PJD007) and the Natural Science Foundation of Shanghai (No. 14ZR1403500).

\end{document}